\documentclass[a4paper,conference]{IEEEtran}

\usepackage{aas_macros}
\usepackage{cite}      

\usepackage{graphicx}  

\usepackage{subfigure} 

\usepackage{amsmath,amssymb}   
\usepackage{fancyheadings}
\def\vg{{\bf v}_{\rm g}}
\def\vga{{\bf v}^{a}_{\rm g}}

\def\vd{{\bf v}_{\rm d}}
\def\rhog{\rho_{\rm g}}
\def\rhoga{\rho_{{\rm g}, a}}
\def\rhogb{\rho_{{\rm g}, b}}
\def\rhod{\rho_{\rm d}}
\def\deltav{\Delta {\bf v}}
\def\ts{t_{\rm s}}
\def\vb{{\bf v}}
\newcommand{\pder}[2]{\frac{\partial #1}{\partial #2}}

\begin{document}

\title{Two phase mixtures in SPH --- A new approach}
\author{\IEEEauthorblockN{Daniel Price}
\IEEEauthorblockA{School of Physics and Astronomy\\
Monash University\\
Victoria 3800, Australia\\
daniel.price@monash.edu}
\and
\IEEEauthorblockN{Guillaume Laibe}
\IEEEauthorblockA{School of Physics and Astronomy\\
University of St. Andrews\\
Fife KY16 9SS, UK\\
guillaume.laibe@gmail.com}
}

\maketitle

\begin{abstract}
  We present a new approach to simulating mixtures of gas and dust in smoothed particle hydrodynamics (SPH). We show how the two-fluid equations can be rewritten to describe a single-fluid `mixture' moving with the barycentric velocity, with each particle carrying a dust fraction. We show how this formulation can be implemented in SPH while preserving the conservation properties (i.e. conservation of mass of each phase, momentum and energy). We also show that the method solves two key issues with the two fluid approach: it avoids over-damping of the mixture when the drag is strong and prevents a problem with dust particles becoming trapped below the resolution of the gas.
  
  We also show how the general one-fluid formulation can be simplified in the limit of strong drag (i.e. small grains) to the usual SPH equations plus a diffusion equation for the evolution of the dust fraction that can be evolved explicitly and does not require any implicit timestepping. We present tests of the simplified formulation showing that it is accurate in the small grain/strong drag limit. We discuss some of the issues we have had to solve while developing this method and finally present a preliminary application to dust settling in protoplanetary discs.
\end{abstract}

\section{Introduction}
\label{sec:intro}
\subsection{Gas/dust mixtures}
Multiphase flows where particulate matter is carried by a fluid are crucial to many problems in astrophysics and engineering. Perhaps the simplest example is that of gas and dust, described by
\begin{align}
\frac{\partial \rho_{\rm g}}{\partial t} + \nabla\cdot( \rho_{\rm g} \vg ) & = 0, \label{eq:rhog} \\
\frac{\partial \rho_{\rm d}}{\partial t} + \nabla\cdot( \rho_{\rm d} \vd ) & = 0, \\
\rhog \left[ \frac{\partial \vg}{\partial t} + (\vg\cdot\nabla) \vg \right] & = -\nabla P + K (\vd - \vg), \\
\rhod \left[ \frac{\partial \vd}{\partial t} + (\vd\cdot\nabla) \vd \right] & = - K (\vd - \vg), \\
\frac{\partial u}{\partial t} + (\vg\cdot\nabla) u& = -\frac{P}{\rhog} (\nabla\cdot \vg) + K (\vg - \vd)^{2}, \label{eq:dudt}
\end{align}
where the subscripts ${\rm g}$ and ${\rm d}$ refer to the gas and dust, respectively, $\rho$ and $\bf v$ refer to density and velocity and $u$ and $P$ are the specific thermal energy and pressure of the gas, respectively. The basic physics is that gas is affected by pressure where gas is not, and that the two fluids are coupled by a drag term (with drag coefficient K derived from the microphysics of the interaction) which is dissipative.  Additional terms arise when the particulate size is finite, giving rise to buoyancy forces. We neglect these since the volume occupied by dust grains is negligibly small in most astrophysical problems. We also neglect the thermal coupling between the phases.

\subsection{The stopping time}
The physics of drag is best quantified by the `stopping time'
\begin{equation}
t_{\rm s} \equiv \frac{\rhog \rhod}{K (\rhog + \rhod)},
\end{equation}
which is the characteristic timescale on which the two fluids dissipate their relative motion and stick together at a common velocity. It is inversely proportional to the strength of the mutual drag force. Short stopping times mean that the mixture moves together at a common velocity (the velocity of the barycentre), long stopping times mean that the two fluids move separately, but intermediate stopping times produce the most dissipation, since drag acts to dissipate the relative motion.

\subsection{Gas/dust mixtures in SPH}
The traditional approach to gas/dust mixtures in Smoothed Particle Hydrodynamics (SPH) is to discretise these equations using a different set of particles for each phase of the mixture \cite{monaghankocharyan95}, giving (\cite{laibeprice12,laibeprice12a})
\begin{align}
\rhog^{a} = & \sum_{b} m_{b} W(\vert {\bf r}_{a} - {\bf r}_{b}\vert, h_{a}), \label{eq:rhogsph}\\
\rhod^{i}  = & \sum_{j} m_{j} W(\vert {\bf r}_{i} - {\bf r}_{j}\vert, h_{i}), \\
\frac{{\rm d}\vga}{{\rm d}t} = & -\sum_{b} m_{b} \left[\frac{P_{a}}{\Omega_{a}\rhoga^{2}} \nabla_{a}W_{ab}(h_{a}) + \frac{P_{a}}{\Omega_{b}\rhogb^{2}} \nabla_{a}W_{ab}(h_{b})\right]  \nonumber \\
& -\nu \sum_{j} m_{j} \frac{{\bf v}_{aj} \cdot \hat{\bf r}_{aj}}{(\rho_{a} + \rho_{j}) t^{\rm s}_{aj}}  D_{aj} (h_{a}), \label{eq:dvg} \\
\frac{{\rm d}\vd^{i}}{{\rm d}t} = & -\nu \sum_{b} m_{b} \frac{{\bf v}_{ib} \cdot \hat{\bf r}_{ib}}{(\rho_{i} + \rho_{b}) t^{\rm s}_{ib}}  D_{ib} (h_{b}), \label{eq:dvd}  \\
\frac{{\rm d}u}{{\rm d}t} = & -\frac{P_{a}}{\rhoga^{2}}  \sum_{b} m_{b} {\bf v}_{ab}\cdot \nabla_{a} W_{ab} (h_{a})\\
& \nu \sum_{j} m_{j} \frac{({\bf v}_{aj} \cdot \hat{\bf r}_{aj})^{2}}{(\rho_{a} + \rho_{j}) t^{\rm s}_{aj}}  D_{aj} (h_{a}), \label{eq:dudtsph}
\end{align}
where we use labels $a$,$b$ and $c$ to refer to gas particles; $i$, $j$ and $k$ to refer to dust particles and $\nu$ is the number of dimensions.

In recent work \cite{laibeprice12,laibeprice12a} we made a number of key improvements to the discretisation of this set of equation in SPH. We found a factor of 10 improvement in accuracy at no additional cost by employing a double-hump shaped kernel ($D_{aj}$ in equations \ref{eq:dvg} and \ref{eq:dvd}), to compute the drag terms instead of the usual bell shaped kernel ($W_{aj}$). We also presented an improved implicit integration method and generalised the earlier methods of \cite{monaghankocharyan95,monaghan97a} to spatially variable smoothing lengths.

\begin{figure}
\begin{center}
\includegraphics[width=\columnwidth]{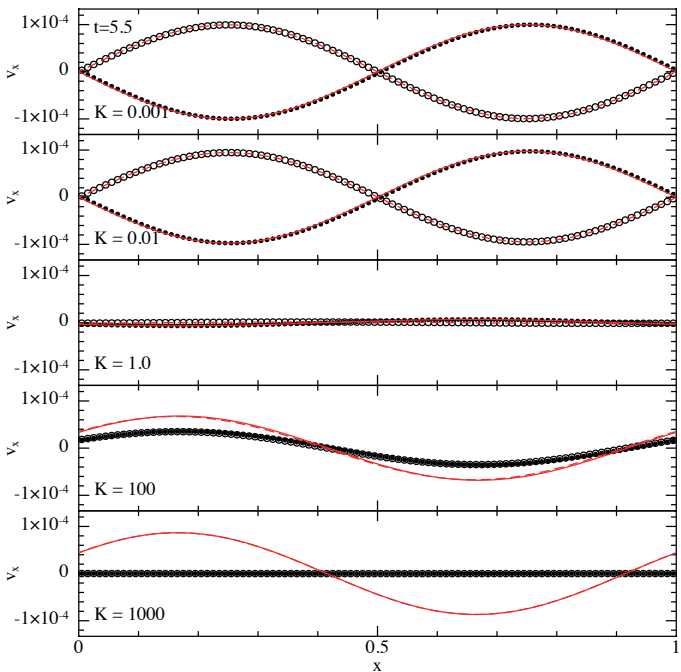}
\caption{Linear waves in a dust-gas mixture (the `{\sc dustywave}' problem), showing the SPH two fluid solution with $2 \times 100$ particles (black circles=gas; open circles=dust) after 5.5 periods, compared to the analytic solution for the gas (solid red line) and dust (dashed red line). At low and intermediate drag the solution is accurate, but at strong drag the numerical solution because the short lengthscale separating the two fluids is not resolved.}
\label{fig:dustywave}
\end{center}
\end{figure}
\subsection{Two problems with two fluids}
\subsubsection{Overdamping}
\label{sec:overdamping}
 The first problem we found with developing algorithms for dust/gas mixtures was that there were few simple test problems which could be used to benchmark the numerical solution. This led one of us (GL) to derive the complete analytic solution for linear waves in the mixture, which we published in \cite{laibeprice11}. We found this immensely useful and enlightening, and indeed it revealed a rather fundamental limitation to the two fluid formulation. Figure~\ref{fig:dustywave} shows a typical SPH two-fluid solution after 5$\frac12$ wave periods, solving Equations~(\ref{eq:rhogsph})--(\ref{eq:dudtsph}) for a one dimensional linear-wave in an equal mixture of dust and gas while varying the drag parameter $K$, in each case compared to the analytic solution in both the gas (solid red line) and dust (dashed red line).
 
  The behaviour of the analytic solution is intuitive --- when the drag is small the solution corresponds to an undamped sound wave propagating in the gas. At very strong drag the solution also corresponds to an undamped soundwave, with the only effect being a change to the effective sound speed because of the weight of the dust being carried along by the gas. Importantly it is only at intermediate drag (stopping times comparable to the wave period) that the solution should be strongly damped.
  
   The SPH solution, by contrast, shows a strongly damped solution at high drag. The reason is that to accurately capture the physics when the drag is strong, one must resolve the (very) short lengthscale separating the two fluids. Indeed, we show in \cite{laibeprice12} that using a very large number of particles (up to 10,000 in 1D) \emph{does} reproduce the analytic solution, but the resolution requirement is prohibitive (corresponding to $h \lesssim t_{\rm s} c_{\rm s}$, where $c_{\rm s}$ is the sound speed). The effect of under-resolving this length scale is to mimic the effect of a much larger separation length, giving a solution closer to that of an intermediate drag which is highly dissipative. This \emph{spatial} resolution requirement is in addition to the usual stability constraint on the timestep $\Delta t < t_{\rm s}$ from the drag terms. In other words, \emph{the two fluid method requires an infinite number of particles and an infinite number of timesteps to correctly resolve the limit of a perfectly coupled mixture}. 

\begin{figure}
\begin{center}
\includegraphics[width=\columnwidth]{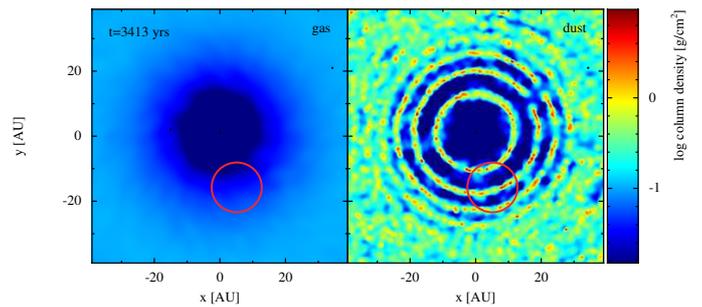}
\caption{Gas (left) and dust (right) column density in a two-fluid simulation of material orbiting in a protoplanetary disc. Once the dust smoothing length becomes smaller than the typical gas smoothing length (solid red circle shows a $2h$ for a representative gas particle) the dust particles become artificial `trapped' in high density rings, due to the lack of mutual repulsion between SPH dust particles.}
\label{fig:dust-trapping}
\end{center}
\end{figure}

\subsubsection{Artificial trapping of dust particles}
 Because they do not feel any mutual repulsion, dust particles can also become artificially `trapped' in high density regions. We found this to occur whenever the dust collects on a scale smaller than the local smoothing length of gas particles. An example is shown in Figure~\ref{fig:dust-trapping}, showing the dust particles in the centre of a protoplanetary disc simulation forming artificial `rings' once the dust smoothing length is smaller than the typical gas resolution (shown by the red circle). This problem could be only partially mitigated by using the maximum smoothing length of the two fluids in the drag terms (Equations~\ref{eq:dvg} and \ref{eq:dvd}) --- the simulation shown in Figure~\ref{fig:dust-trapping} does this but we still found particle trapping to occur.
 
 Both of these issues motivated us to develop an alternative approach that correctly captures the limit of strong drag/short stopping time, which is the subject of the present contribution.

\section{Dusty gas with one fluid}
\label{sec:onef}
 From the {\sc dustywave} analytic solution one can intuitively understand the physics of the mixture in the limit of infinite drag/short stopping time. This is the limit in which the fluid behaves as a single fluid mixture with a modified sound speed. This motivated us to rewrite the mathematical equations in such a way that the limit of $t_{\rm s} \to 0$ is obvious from the mathematics.
 
\subsection{One fluid to rule them all}
To do this, we simply employed a change of variables. Instead of solving for $\rhod$, $\rhog$, $\vg$ and $\vd$, in \cite{laibeprice14} we rewrote the equations in terms of the total density $\rho$, the dust fraction $\epsilon \equiv \rhod/\rho$, the barycentric velocity ${\bf v}$ and the relative velocity between the fluids $\Delta {\bf v}$, where
\begin{equation}
{\bf v} \equiv \frac{\rhog\vg + \rhod\vd}{\rhog + \rhod},
\end{equation}
and
\begin{equation}
\Delta {\bf v} \equiv \vd - \vg.
\end{equation}
The original variables can be written in terms of the new variables according to
\begin{align}
\rhog & = ( 1 - \epsilon) \rho, \label{eq:brhog} \\
\rhod & = \epsilon \rho,  \label{eq:brhod} \\
\vg & = {\bf v} - \epsilon \deltav, \label{eq:bvg} \\
\vd & = {\bf v} + ( 1 - \epsilon) \deltav. \label{eq:bvd}
\end{align}

 The evolution equations in the new variables, written in Lagrangian form with the time derivative defined for a single fluid moving with the barycentric velocity,
\begin{equation}
 \frac{\rm d}{{\rm d} t} \equiv \pder{}{t} + \left({\bf v}\cdot\nabla\right),
\end{equation}
are given by
\begin{align}
\frac{{\rm d} \rho}{{\rm d} t} = & - \rho (\nabla \cdot {\bf v}), \label{eq:genmass_rho} \\
\frac{{\rm d} \epsilon}{{\rm d} t} = & -\frac{1}{\rho} \nabla \cdot \left[ \epsilon\left(1 - \epsilon \right) \rho \deltav \right] \label{eq:gendtgevol} , \\
\frac{{\rm d} {\bf v}}{{\rm d} t} = & - \frac{\nabla P}{\rho} - \frac{1}{\rho}\nabla\cdot \left[ \epsilon\left(1 - \epsilon \right) \rho \deltav \deltav \right] + \mathbf{f}\label{eq:genmomentum_bary},\\
\frac{{\rm d} \deltav}{{\rm d} t} = &  - \frac{\deltav}{t_{\rm s}} + \frac{\nabla P}{\rhog} - (\deltav \cdot \nabla) {\bf v} \nonumber \\
& + \frac{1}{2}\nabla \left[ \left(2 \epsilon - 1 \right) \deltav ^{2} \right]. \label{eq:genmomentum_deltav} \\
\frac{{\rm d} u}{{\rm d} t} = &  -\frac{P_{\mathrm{g}}}{\rhog} \left(\nabla \cdot \vg \right)  +   \epsilon \left( \deltav \cdot \nabla\right) u  + \frac{\epsilon\deltav^{2}}{t_{\rm s}}, \label{eq:dudtonef}
\end{align}
where ${\bf f}$ represents body forces such as a gravitational potential. This change of variables entails no loss of information but reveals a great deal of the physics.

\subsection{Physical interpretation}
First, we see that in the limit of $\deltav \to 0$ and $\epsilon \to 0$ the equations reduce to the usual equations of fluid dynamics, which we know can be solved using standard SPH techniques. Second, we see the physics of the drag clearly in Equation~\ref{eq:genmomentum_deltav} --- the pressure gradient acts as the `source term' to grow the differential velocity, while the stopping time 
appears as the characteristic decay timescale. That is, in the absence of pressure gradients the relative velocity decays exponentially to zero on the stopping time. Furthermore terms such as the anisotropic pressure in Equation~\ref{eq:genmomentum_bary} that are second order in $\deltav$ can be neglected when the differential velocity is small, as occurs when the drag is strong and the stopping time is short. So it is clear that a discretisation based on Equations~\ref{eq:genmass_rho}--\ref{eq:dudtonef} will reduce to a standard SPH formulation in the previously `impossible' limit of $\ts \to 0$.

\begin{figure}
\begin{center}
\includegraphics[width=\columnwidth]{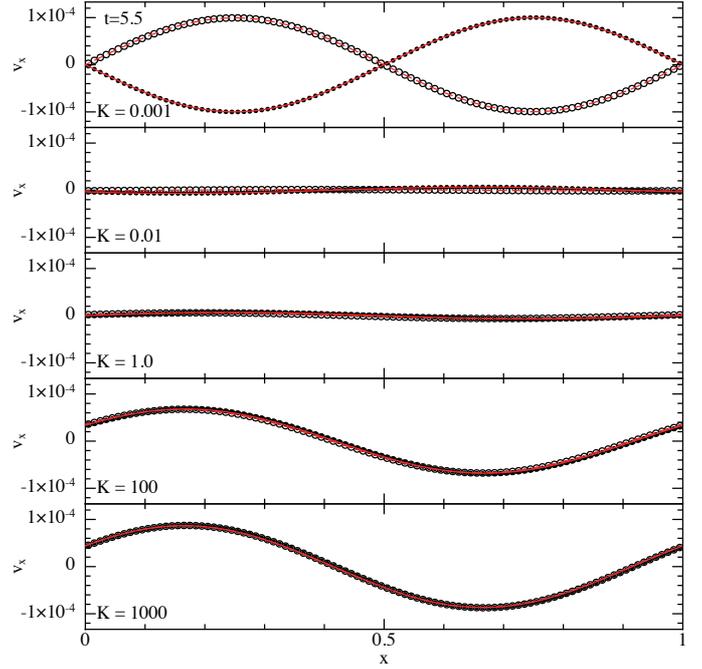}
\caption{Solution to the {\sc dustywave} problem using the general one fluid SPH method, showing gas and dust properties reconstructed from the SPH mixture particles (black points and circles) compared to the analytic solution (red solid and dashed lines for gas and dust, respectively). The SPH solution is in excellent agreement with the analytic solution in all cases, unlike for the two fluid method (Figure~\ref{fig:dustywave}).}
\label{fig:dustywave-onef}
\end{center}
\end{figure}

\subsection{SPH discretisation of the general one fluid formulation}
In principle there is no loss of information in solving Equations~\ref{eq:genmass_rho}--\ref{eq:dudtonef} instead of \ref{eq:rhog}--\ref{eq:dudt} but the two approaches are conceptually very different. The key difference is that we must now consider only \emph{one} set of particles moving with the barycentre of the fluid, advected according to
\begin{equation}
\frac{{\rm d}{\bf x}}{{\rm d}t} = {\bf v}.
\end{equation}
 These particles represent neither gas nor dust, but rather the `mixture'. The dust fraction $\epsilon$ and likewise the differential velocity $\deltav$ are now intrinsic properties of the mixture that are advected along with the particles. The evolution of $u$ is modified to express the fact that the particles are no longer simply gas particles. One must also ensure that the conservation laws are satisfied --- in particular that the mass of each phase is separately conserved, just as it would be with the two fluid approach.

 Nevertheless it is reasonably straightforward to write down the SPH discretisation in a way that does satisfy the conservation properties. The resulting equations, ignoring for the moment artificial dissipation terms, are given by \cite{laibeprice14a}
 \begin{align}
\rho_{a} = & \sum_{b} m_{b} W_{ab} (h_{a}), \label{eq:density-final} \\
\frac{{\rm d}\epsilon_{a}}{{\rm d} t} = & - \sum_{b} m_{b} \left[ \frac{\epsilon_{a} \left(1 - \epsilon_{a} \right)}{\Omega_{a} \rho_{a}}  \deltav_{a}\cdot\nabla_{a} W_{ab} (h_{a})\right. \nonumber \\
 & \hspace{1.2cm} + \left. \frac{\epsilon_{b} \left(1 - \epsilon_{b} \right)}{\Omega_{b} \rho_{b}}\deltav_{b}\cdot \nabla_{a} W_{ab} (h_{b}) \right], \label{eq:dusttogas-final} \\
\frac{{\rm d} \vb_{a}}{{\rm d} t} = & -\sum_{b} m_{b} \left[ \frac{P_{a}}{\Omega_{a} \rho_{a}^{2}} \nabla W_{ab}(h_{a}) + \frac{P_{b}}{\Omega_{b} \rho_{b}^{2}} \nabla W_{ab}(h_{b}) \right] \nonumber\\
& -\sum_{b} m_{b} \left[ \frac{\epsilon_{a}\left(1 - \epsilon_{a} \right) \deltav_{a}}{\Omega_{a} \rho_{a}} \deltav_{a}.\nabla W_{ab}(h_{a}) \right. \nonumber \\
&  \phantom{\sum_{b} m_{b}[} + \left.\frac{\epsilon_{b}\left(1 - \epsilon_{b} \right)\deltav_{b}}{\Omega_{b} \rho_{b}} \deltav_{b}\cdot\nabla W_{ab}(h_{b}) \right] \nonumber \\
& + {\bf f}_{a},
\label{eq:mom-final} \\
\frac{{\rm d}{\deltav}_{a}}{{\rm d}t} = &  -\frac{\deltav_{a}}{t_{\rm s, a}} \nonumber \\
- & \frac{\rho_{a}}{\rhoga}\sum_{b} m_{b} \left[ \frac{P_{a}}{\Omega_{a} \rho_{a}^{2}} \nabla W_{ab}(h_{a}) + \frac{P_{b}}{\Omega_{b} \rho_{b}^{2}} \nabla W_{ab}(h_{b}) \right], \nonumber \\
+ &\frac{1}{\rho_{a}\Omega_{a}}\sum_{b} m_{b} ({\bf v}_{a} - {\bf v}_{b}) \deltav_{a}.\nabla W_{ab} (h_{a}) \nonumber \\
+ & \frac{1}{2\rho_{a}\Omega_{a}} \sum_{b} m_{b} \left[ \left(1 - 2 \epsilon_{a} \right) \deltav_{a}^{2} \right.\nonumber \\
& \hspace{1.8cm} -\left.\left(1 - 2 \epsilon_{b} \right) \deltav_{b}^{2} \right] \nabla W_{ab} (h_{a}), \label{eq:deltav-final}\\ 
\frac{{\rm d}u_{a}}{{\rm d}t} = & \frac{P_{a}}{\Omega_{a} \rho_{a} \rho_{{\rm g},a}} \sum_{b} m_{b} \left({\bf v}_{{\rm g},a} - {\bf v}_{{\rm g},b} \right) \cdot \nabla W_{ab}(h_{a}) \nonumber \\
& - \frac{\epsilon_a}{\Omega_{a} \rho_{a}}  \sum_{b} m_{b} (u_{a} - u_{b}) \deltav_{a} .\nabla W_{ab} (h_{a}) \nonumber \\
& + \epsilon_a \frac{\deltav_{a}^{2}}{t_{\rm s, a}}. \label{eq:dudt-final}
\end{align}
The full set of equations including the modifications to artificial viscosity necessary for shock capturing are given in \cite{laibeprice14a}.

\subsection{Tests of the general one fluid method}
Figure~\ref{fig:dustywave-onef} shows the solution obtained on the {\sc dustywave} problem by solving Equations \ref{eq:density-final}--\ref{eq:dudt-final} instead of \ref{eq:rhogsph}--\ref{eq:dudtsph}. One can see immediately that there is no overdamping of the fluid when the drag is strong. Furthermore, the analytic solution is now correctly reproduced in all cases, including the limits of both strong and weak drag (i.e. for both $\ts \to \infty$ and $\ts \to 0$). It is intuitive why this should be the case, since the resolution problem with the two fluid method referred to in Section~\ref{sec:overdamping} was related to the need to resolve the physical separation length between gas and dust particles. In the one fluid method there is no physical `separation' of the `gas' particles from the `dust' particles --- every particle carries information about both phases. (Very!) careful inspection of Figure~\ref{fig:dustywave-onef} would reveal this, since to visualise the gas and dust `particles' we have simply made two copies of the same set of mixture particles, one with the gas properties reconstructed using (\ref{eq:brhog}) and (\ref{eq:bvg}) and one with the dust properties reconstructed using (\ref{eq:brhod}) and (\ref{eq:bvd}). 

\begin{figure}
\begin{center}
\includegraphics[width=\columnwidth]{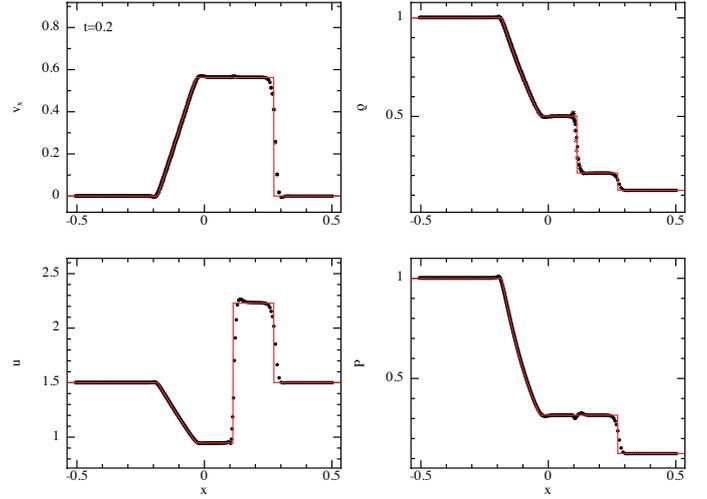}
\caption{Solution to a dust-gas shock problem with strong drag, showing the solution with the one fluid method (black points) compared to the analytic solution in the limit where the stopping time is small (red line).}
\label{fig:dustyshock}
\end{center}
\end{figure}

A further confirmation that the one fluid method solves the resolution issue is shown in Figure~\ref{fig:dustyshock}, showing the propagation of a shock in the mixture when the drag is strong. The solution in this case should be identical to a hydrodynamic shock but with the shock propagating at the modified sound speed due to the weight of the dust. It can be seen that the one fluid method gives results in excellent agreement with the analytic solution (red line), whereas in \cite{laibeprice12} we found that around 10,000 particles in one dimension were needed to produce the correct solution with the two fluid method.

 It is also intuitively obvious that the one fluid approach also solves the dust trapping problem --- there is no separate set of `dust' particles, so they cannot become trapped below the resolution of the gas. Instead, the resolution of both phases is now tied to the density of the mixture rather than being separate for each phase. Thus \emph{by definition} the gas and the dust components are resolved with the same resolution (smoothing) length.

\subsection{Limitations of the general one fluid approach}
Despite the general nature of the one fluid formulation and the removal of the spatial resolution criterion at high drag, there remains the stability constraint on the timestep from the stopping time if explicit timestepping is used to solve (\ref{eq:density-final})--(\ref{eq:dudt-final}), i.e.
\begin{equation}
\Delta t < t_{\rm s}.
\end{equation}
As discussed in \cite{laibeprice14}, implementing implicit timestepping to solve Equations~\ref{eq:deltav-final} and \ref{eq:dudt-final} when $\ts$ is small is considerably simpler than implementing an implicit method for the two fluid method (\cite{monaghan97a,laibeprice12a,lorenbate14}). Nevertheless it represents an extra complication to the algorithm.

A further limitation is that the single fluid description is a \emph{fluid} description. In particular the velocity field is assumed to be single-valued. This assumption breaks down in the limit where there is little or no drag ($\ts \to \infty$) and the dust particles act as a pressure-less independent set of particles, uncoupled from the gas. For astrophysics this occurs for large km-sized planetesimals in protoplanetary discs, which oscillate freely around the midplane and form structures which are supported by the velocity dispersion of their constituent particles. The velocity field in this case can be multi-valued which can be captured when representing the dust as a separate set of particles but not if represented as part of a mixture. 

This means that the one fluid formulation is mainly useful in the limit where the stopping time is short compared to other timescales in the simulation, which is the limit of small (cm sized and smaller in protoplanetary discs) grains in astrophysics. However, in this limit we can also derive a much simpler method.

\section{The diffusion approximation for the mixture}

\subsection{The terminal velocity approximation}
 In the limit where $\ts$ is smaller than the computational timestep, Equations~\ref{eq:genmass_rho}--\ref{eq:dudtonef} can be simplified by neglecting terms that are second order in $\deltav$ \cite{laibeprice14}. This is known as the `terminal velocity' approximation (e.g. \cite{youdingoodman05,chiang08}), since the time-dependence in $\deltav$ can be neglected and for the case of hydrodynamics we simply have
\begin{equation}
\deltav \approx \ts \frac{\nabla P}{\rhog}, \label{eq:tv}
\end{equation}
with (\ref{eq:genmass_rho})--(\ref{eq:dudtonef}) simplifying to (\cite{laibeprice14,pricelaibe15})
\begin{align}
\frac{\mathrm{d} \rho}{\mathrm{d} t} = & -\rho (\nabla\cdot\vb) \label{eq:ctyalt},\\
\frac{\mathrm{d} \vb}{\mathrm{d}t}  = & -\frac{\nabla P}{\rho}  + \mathbf{f} \label{eq:momalt}, \\
\frac{\mathrm{d}\epsilon}{\mathrm{d}t} = & -\frac{1}{\rho} \nabla\cdot\left(\epsilon \ts \nabla P  \right), \label{eq:epsalt} \\
\frac{{\rm d} u}{{\rm d} t} = &  -\frac{P}{\rhog} (\nabla\cdot\vb) - \frac{\epsilon \ts}{\rhog} \left(  \nabla P \cdot\nabla u \right). \label{eq:dudtalt}
\end{align}
In this limit we simply recover the usual equations of hydrodynamics supplemented by an evolution equation (\ref{eq:epsalt}) for the dust fraction and an additional term in the energy equation (\ref{eq:dudtalt}). The pressure is also modified by the presence of the dust since it depends on the gas density rather than the total density, giving for example
\begin{equation}
P = (\gamma - 1) \rhog u = (1-\epsilon) (\gamma - 1) \rho u.
\end{equation}
The above change to the pressure, and hence the sound speed, produces the zeroth order effect of a `heavy fluid' seen in the analytic solution for the {\sc dustywave} (Figs.~\ref{fig:dustywave} and \ref{fig:dustywave-onef}). 

 If desired, one can simplify the formulation even further by making the approximation that $\ts = 0$. This leaves only the `heavy fluid' effect, giving precisely the limit of sound waves propagating at the modified sound speed. This limit would also imply --- from Equation~\ref{eq:epsalt} --- a constant dust-to-gas ratio, which is what is commonly assumed when interpreting most astronomical observations of dust in space.
 
\subsection{SPH discretisation}
The SPH discretisation is also much simpler. Both (\ref{eq:ctyalt}) and (\ref{eq:momalt}) are identical to the usual hydrodynamic equations, so the only question is how to discretise (\ref{eq:epsalt}) and the extra term in (\ref{eq:dudtalt}). There are two possible ways of discretising these terms; either make the terminal velocity approximation (\ref{eq:tv}) in (\ref{eq:density-final})--(\ref{eq:dudt-final}), or simply discretise the terms in (\ref{eq:epsalt}) and (\ref{eq:dudtalt}) directly. We compare both approaches in \cite{pricelaibe15}, but find only minor differences, and so here present only the `direct second derivatives' version since it is both simpler and computationally more efficient.

 Adopting a standard approach to writing second derivatives in SPH \cite{clearymonaghan99,espanolrevenga03,price12} we discretise (\ref{eq:ctyalt})--(\ref{eq:dudtalt}) according to
\begin{align}
\rho_{a} = & \sum_{b} m_{b} W_{ab} (h_{a}), \label{eq:sphcty} \\
\frac{{\rm d}{\bf v}_a}{{\rm d}t} = & -\sum_{b} m_{b} \left[ \frac{P_{a}}{\Omega_{a} \rho_{a}^{2}} \nabla_{a} W_{ab} (h_{a}) + \frac{P_{b}}{\Omega_{b} \rho_{b}^{2}} \nabla_{a} W_{ab} (h_{b}) \right] \nonumber \\
& + {\bf f}_{a}, \label{eq:sphmom} \\
\frac{\mathrm{d}\epsilon_{a}}{\mathrm{d}t} = & - \sum_{b} \frac{m_{b}}{\rho_{a}\rho_{b}} (D_{a} + D_{b}) \left(P_{a} - P_{b} \right)\overline{F}_{ab}, \label{eq:sphepsalt}\\
\frac{{\rm d}u_{a}}{{\rm d}t} = & \frac{P_{a}}{\Omega_{a} (1 - \epsilon_{a})\rho_{a}^{2}} \sum_{b} m_{b} {\bf v}_{ab} \cdot\nabla_{a} W_{ab} (h_{a}) \nonumber \\
 + & \frac{1}{2(1-\epsilon_{a})\rho_{a}}\sum_{b} \frac{m_{b}}{\rho_{b}} (u_{a} - u_{b}) (D_{a} + D_{b}) (P_{a} - P_{b}) \overline{F}_{ab}, \label{eq:sphdudtalt}
\end{align}
where $D_{a} \equiv  \epsilon_{a} \ts^{a}$ and the scalar kernel function $F_{ab}$ is defined in the usual manner such that $\nabla W_{ab} \equiv F_{ab} {\bf r}_{ab}$ and $\overline{F}_{ab} = \frac12 [F_{ab}(h_{a}) + F_{ab}(h_{b})]$. 

The second term in Equation~\ref{eq:sphdudtalt} is one of the most bizarre SPH discretisations we have ever come across, but derives from the discretisation of (\ref{eq:sphepsalt}) combined with the need to conserve energy. A remarkable proof is given in the appendix of \cite{pricelaibe15} that this is indeed a correct discretisation of the corresponding term in (\ref{eq:dudtalt}).

As with the general one-fluid method, all of the conservation properties of the two fluid method are satisfied by (\ref{eq:sphcty})--(\ref{eq:sphdudtalt}). That is, the mass of each phase as well as the total mass, momentum, angular momentum and energy are all conserved exactly.

\subsection{Timestep constraints in the diffusion approximation}
In a medium with a uniform density and temperature (such that $P = c_{\rm s}^{2} \rhog = c_{\rm s}^{2} (1 - \epsilon) \rho$), Equation~\ref{eq:epsalt} becomes a simple diffusion equation for the dust fraction
\begin{equation}
\frac{{\rm d}\epsilon}{{\rm d}t} = \nabla\cdot\left(\eta \nabla \epsilon  \right), \label{eq:diffdust}
\end{equation}
where the diffusion coefficient is given by
\begin{equation}
\eta \equiv \epsilon \ts c_{\rm s}^{2}.
\end{equation}
This implies a stability constraint of the form
\begin{equation}
\Delta t < \frac{h^{2}}{\epsilon \ts c_{\rm s}^{2}}.
\end{equation}
This constraint, while second order in the resolution length, presents the \emph{opposite} constraint on the timestep compared to both the two fluid method (Section~\ref{sec:intro}) and the general one fluid method (Section~\ref{sec:onef}). That is, the timestep is \emph{inversely} proportional to the stopping time, whereas with the other approaches the timestep is \emph{proportional} to $\ts$.

 This implies that Equations (\ref{eq:sphcty})--(\ref{eq:sphdudtalt}) require only explicit timestepping to correctly capture the limit of small grains/short stopping time/strong drag --- a stark contrast to the situation with the two fluid method.

\begin{figure}
\begin{center}
\includegraphics[width=\columnwidth]{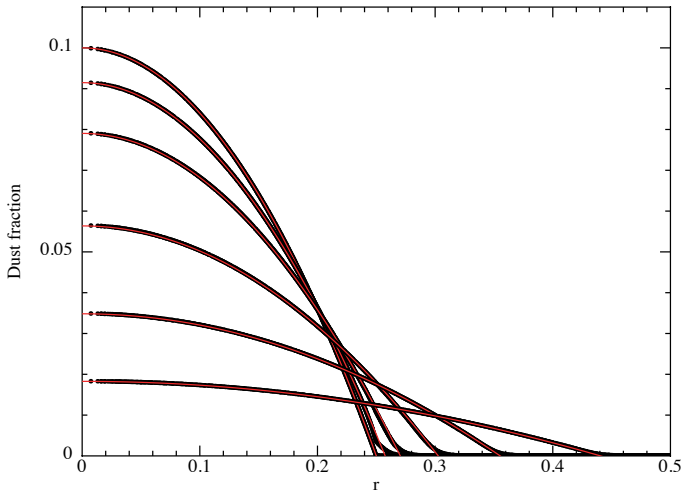}
\caption{Test problem showing diffusion of the dust fraction in a three dimensional calculation. The plot shows the SPH solution of the diffusion equation (\ref{eq:diffdust}) at various times, compared to the analytic solution(s) given by the solid red lines.}
\label{fig:diffuse}
\end{center}
\end{figure}

\subsection{Tests of the diffusion method}
\subsubsection{Diffusion of the dust fraction}
The nature of the approximation suggests a simple test problem where the particles positions are fixed but the dust fraction is allowed to evolve according to (\ref{eq:epsalt}). Assuming a pressure proportional to the gas density such that the evolution reduces to (\ref{eq:diffdust}), the analytic solution is well known. and hence can be used. Although the problem is not particularly physical (since the other fluid properties are held fixed) it is useful to benchmark the discretisation of the diffusion equation. 

\begin{figure}
\begin{center}
\includegraphics[width=\columnwidth]{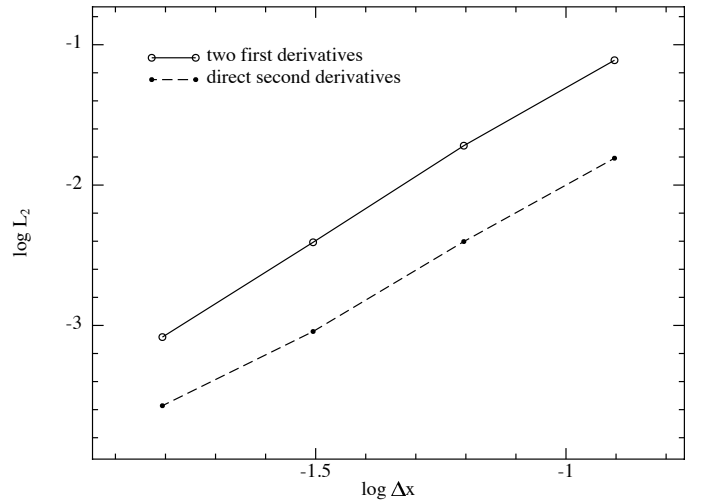}
\caption{Convergence study on the dust diffusion problem, showing second order accuracy of the method in $L_{2}$.}
\label{fig:conv}
\end{center}
\end{figure}

Figure~\ref{fig:diffuse} shows the results of a three dimensional simulation employing $50^{3}$ particles where the dust fraction was initially set to
\begin{equation}
\epsilon(r,0) = \epsilon_{0}\left[1 - \left(\frac{r}{r_{c}}\right)^{2}\right],
\end{equation}
with the numerical solution compared to the analytic solution at various times, assuming a constant $\ts = 0.1$ and density and sound speed of unity. The evolution is indistinguishable from the analytic solution. This is quantified in Figure~\ref{fig:conv} which shows the $L_{2}$ error between the numerical and analytic solutions as a function of resolution (particle spacing) for two different discretisations. As expected, the convergence is second order in the particle separation since the particles remain in an ordered configuration.

\begin{figure}
\begin{center}
\includegraphics[width=\columnwidth]{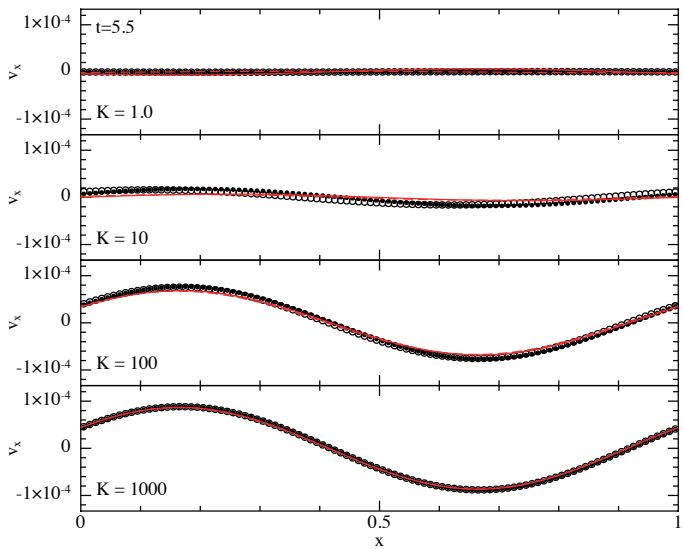}
\caption{SPH simulation of the {\sc dustywave} problem using the diffusion approximation, compared to the analytic solutions given by the red solid and dashed lines (c.f. Figures~\ref{fig:dustywave} and \ref{fig:dustywave-onef}). The diffusion approximation assumes that the stopping time is short compared to the timestep and is therefore only applicable when the drag is strong. However, it requires only explicit timestepping to correctly capture the limit of small grains.}
\label{fig:dustywave-diff}
\end{center}
\end{figure}

\subsubsection{Waves}
The numerical results using our SPH implementation of the diffusion method on the {\sc dustywave} problem are shown in Figure~\ref{fig:dustywave-diff}, compared to the analytic solution given by the solid line. The damping in the mixture is correctly captured both when the drag is strong and even in the regime where the diffusion approximation is no longer valid. The small phase error in the solution with $K=10$ is not from the method itself, but from an inconsistency in the initial conditions used to set up the problem (we start with $\deltav = 0$ but the diffusion approximation assumes that $\deltav$ is finite at all times). 

\subsubsection{Shocks} 
The results on the {\rm dustyshock} problem obtained by solving (\ref{eq:sphcty})--(\ref{eq:sphdudtalt}) are indistinguishable from that obtained with the general one fluid method (Figure~\ref{fig:dustyshock}), with the advantage that only explicit timestepping is required.

\begin{figure*}
\begin{center}
\includegraphics[width=\columnwidth]{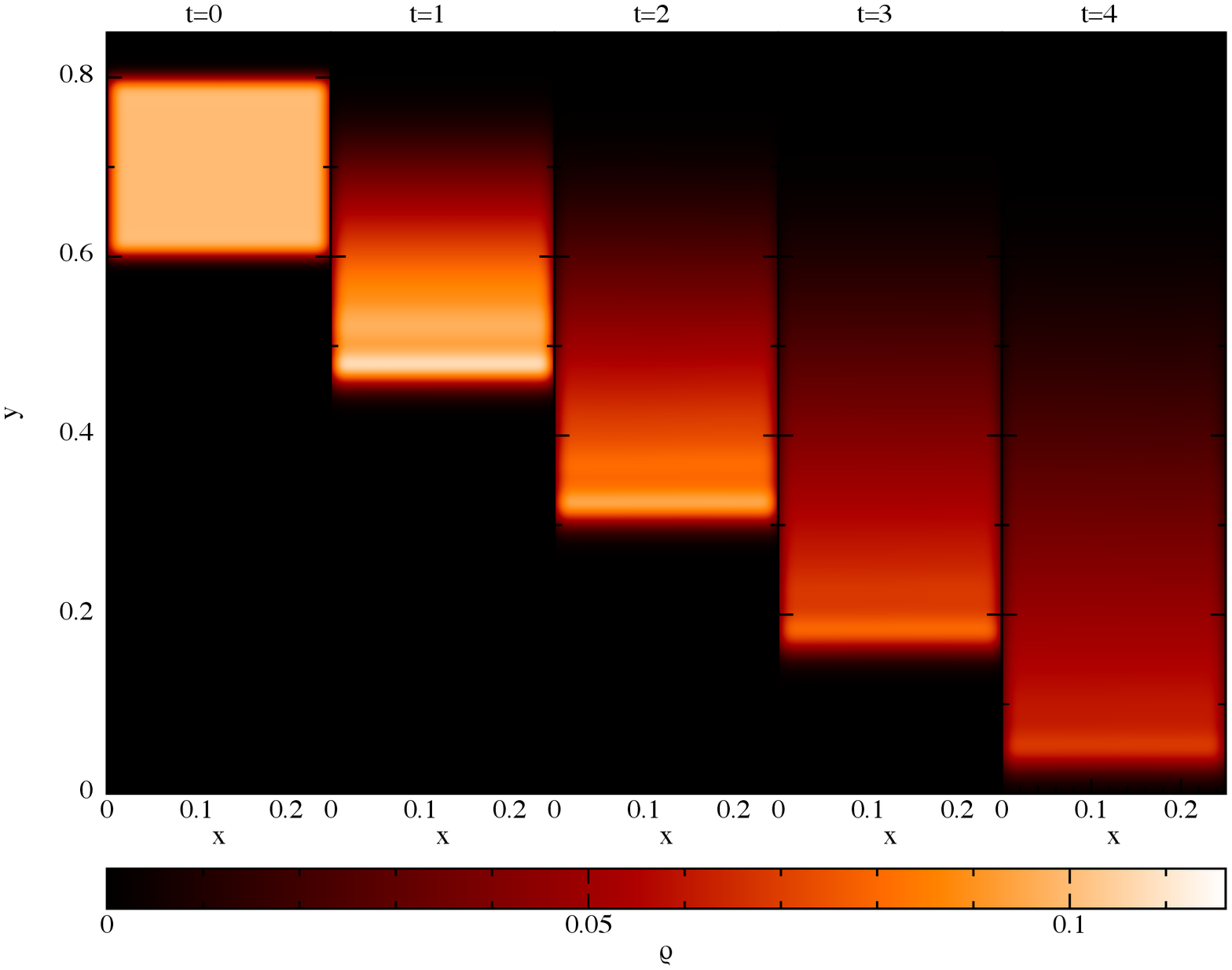}
\includegraphics[width=\columnwidth]{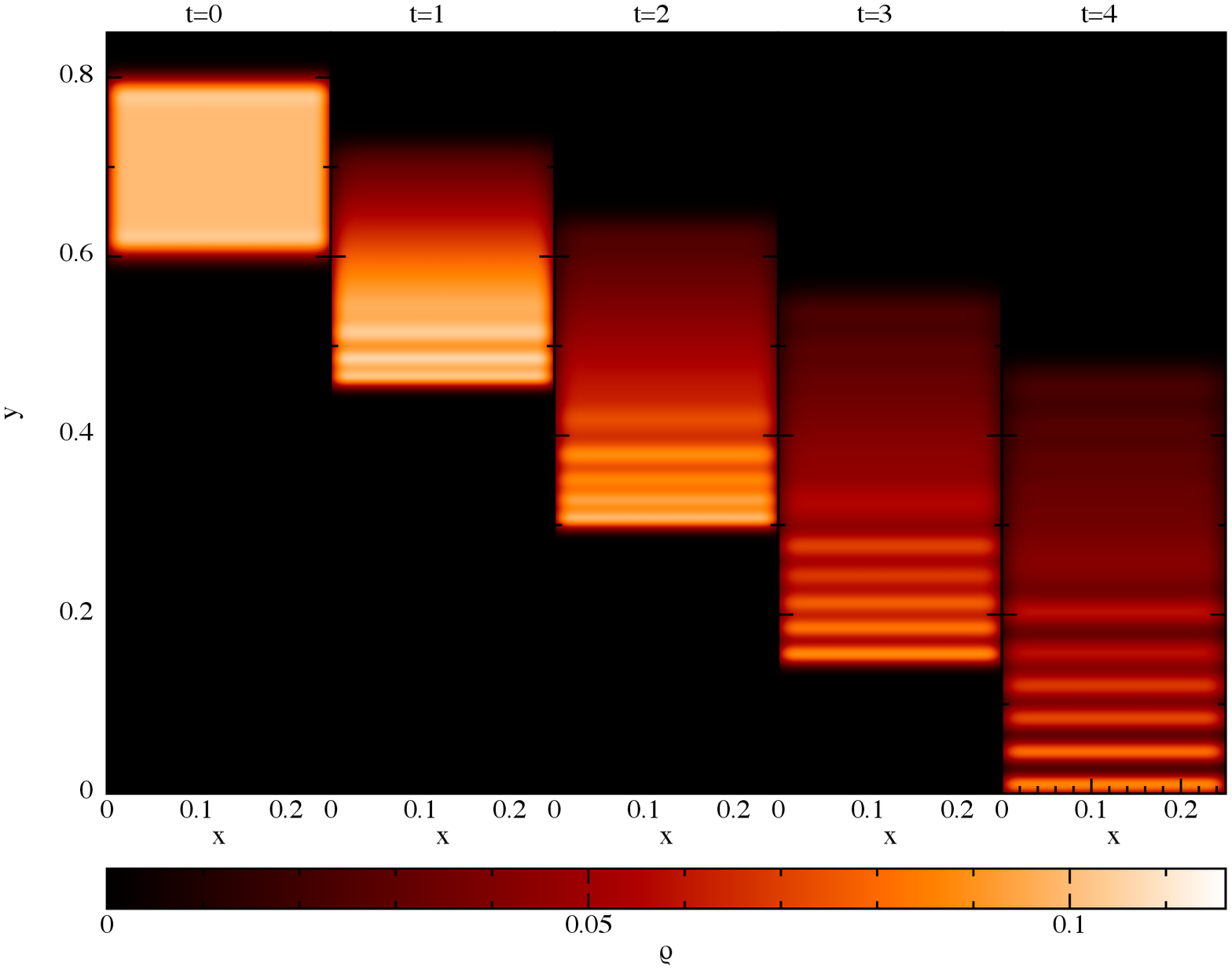}
\caption{Fall of a layer of dust in a stratified atmosphere, comparing results with the diffusion method (left) to the two fluid approach (right).}
\label{fig:dust-sediment}
\end{center}
\end{figure*}
\subsubsection{Fall of a layer of dust}
 A simple problem to illustrate that the diffusion method correctly captures the behaviour of the dust is the `fall of a layer of dust' problem introduced by Monaghan \cite{monaghan97a}. The setup is a box with a vertical gravitational force, with the gas density stratified to balance the vertical gravity and periodic boundaries in the horizontal direction. We use a constant drag coefficient $K=10$, an isothermal equation of state $P = c_{\rm s}^{2} \rho_{\rm g}$ with $c_{\rm s} = 10$ and a gas density $\rhog = 1$ at $y=0$. We set the density stratification, and in the one fluid method the dust location, by slightly altering the particle mass. The dust is initially confined in $0.6 < y < 0.8$ with $\rhod = 0.1$ in the layer and $0$ elsewhere.
 
 Figure~\ref{fig:dust-sediment} compares the results using the diffusion approximation (left) to the results obtained with the standard two fluid method (right). The one fluid method benefits from the regularisation of the mixture particles by the gas pressure, whereas discreteness effects are visible in the two fluid method because the dust particles have no self-interaction. Nevertheless, comparable solutions are obtained with both methods.

\subsubsection{Settling of dust in a protoplanetary disc}
 Our final test problem is drawn from our intended application, namely the settling and migration of dust in discs around young stars during the planet formation process. We consider a vertical section of disc at a particular radius, with the gas pressure in hydrostatic equilibrium with the vertical component of the gravity from the central star. We perform the test at 50 AU with mm-sized dust grains using an Epstein drag prescription (for full details see \cite{pricelaibe15}). The stopping time for grains of this size is a few percent of the orbital timescale, meaning that the terminal velocity approximation is valid. However the problem is still (just) tractable with the two fluid method enabling us to compare all three approaches.
 
\begin{figure}
\begin{center}
\vspace{5mm}
\includegraphics[width=\columnwidth]{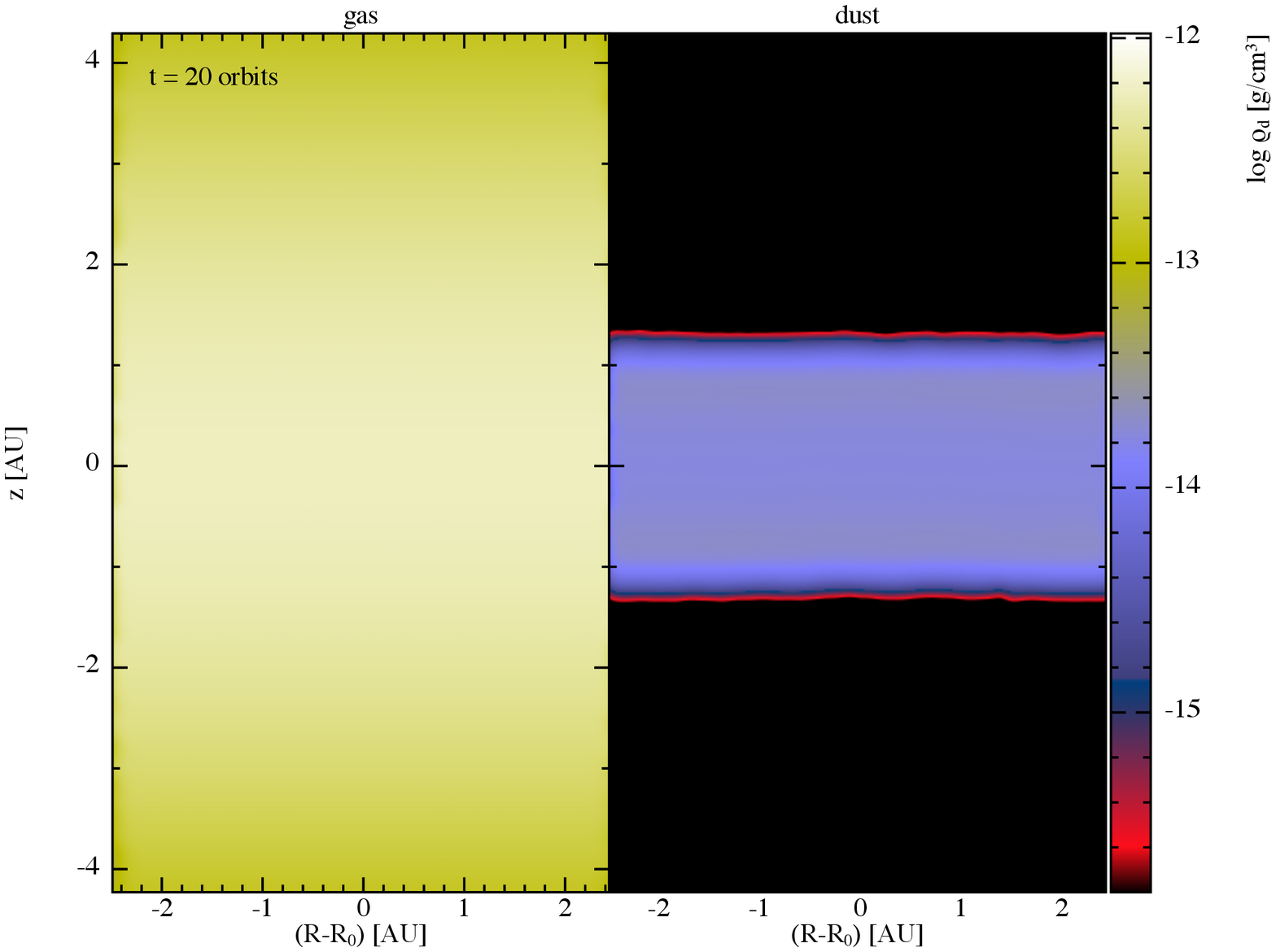}
\includegraphics[width=\columnwidth]{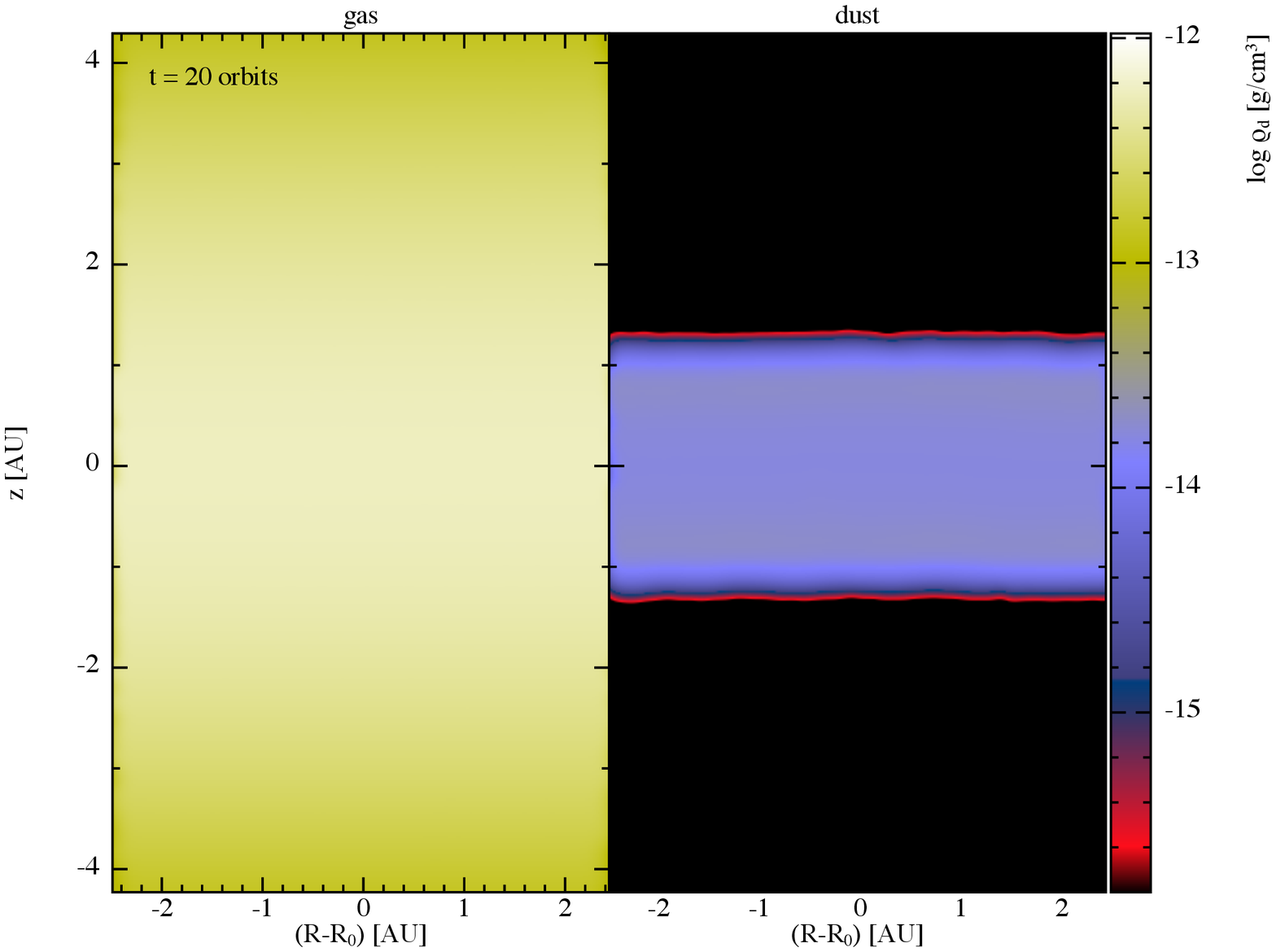}
\vspace{5mm}
\includegraphics[width=\columnwidth]{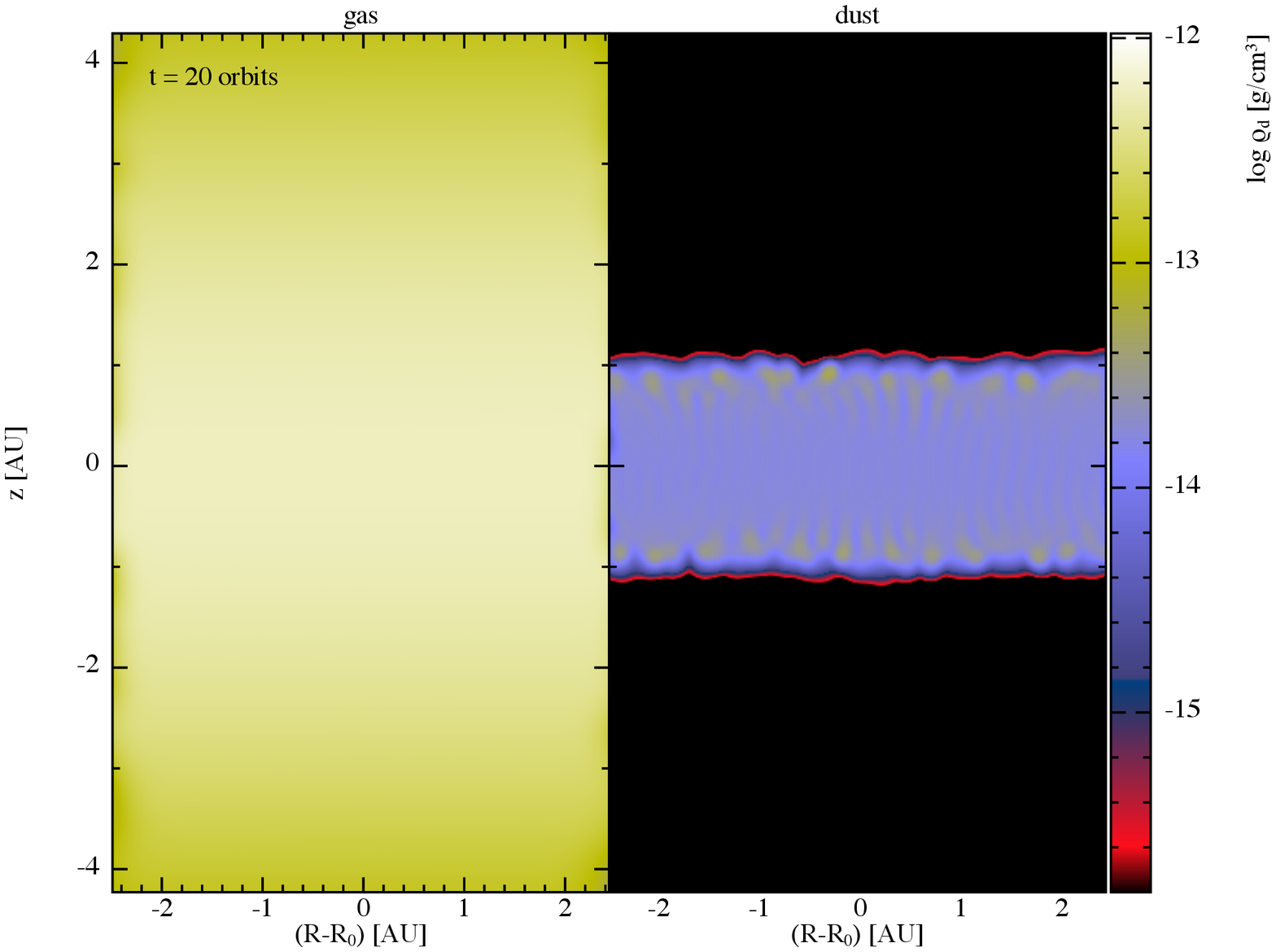}
\caption{Settling of mm-sized dust grains in a vertical section of a protoplanetary disc, showing gas density (left) and dust density (right) computed after 20 orbits using the diffusion approximation (top), the full one fluid method (centre) and the two fluid method (bottom). The solution with the diffusion method is comparable to the other two approaches but is approximately 50 times faster than the two fluid method to compute since it requires only explicit timestepping.}
\label{fig:dust-settle}
\end{center}
\end{figure}

 Figure~\ref{fig:dust-settle} compares the numerical results with the diffusion method (top) with the general one fluid method (centre) and the two fluid method (bottom). The left panel shows the gas density while the right panel shows the dust density after 20 orbits. Settling of dust to the midplane is expected to occur on a timescale of $\sim 10^{2}$ orbits. The dust density is better resolved with the two fluid method because the resolution in the dust follows the dust mass (in contrast to being tied to the total mass in the one fluid case) but it is also more noisy because there is no pressure force to regularise the dust particle distribution. Nevertheless the results demonstrate that the basic physics can be captured with any of the three approaches we have discussed in the paper.
 
 The solution with the diffusion method is $\sim$50 times faster to compute, since it requires only explicit timestepping.

\section{Conclusion}
 We have derived a completely new approach to simulating two-phase mixtures in SPH. Rather than modelling the mixture with two separate sets of particles, we have shown how the equations can be rewritten to describe a single fluid mixture. This solves two key problems with the two fluid approach in the limit where the drag stopping time is short compared to the computational timestep. Furthermore the equations in this limit reduce to simply the usual SPH equations supplemented by a diffusion equation for the dust fraction and a modification to the energy equation and require only explicit timestepping in the limit of small grains. We have so far applied this to dust settling in protoplanetary discs but expect the method to be equally applicable for industrial and engineering problems where particulate matter is carried by a fluid.

\section*{Acknowledgment}
We thank Kieran Hirsh for allowing us to use one of his simulations to create Figure~\ref{fig:dust-trapping}.

\bibliographystyle{IEEEtran.bst}
\bibliography{dan}

\end{document}